# An Efficient Approach towards Mitigating Soft Errors Risks

[1]Muhammad Sheikh Sadi, [1]Md. Mizanur Rahman Khan, [1]Md. Nazim Uddin, [2]Jan Jürjens

[1]Khulna University of Engineering and Technology (KUET), Khulna, Bangladesh,
[2]TU Dortmund, Germany

```
sheikhsadi@gmail.com, rk_mizan@yahoo.com
```

## ABSTRACT

*Smaller feature size, higher clock frequency and lower power consumption are of core concerns of today's nano-technology, which has been resulted by continuous downscaling of CMOS technologies. The resultant 'device shrinking' reduces the soft error tolerance of the VLSI circuits, as very little energy is needed to change their states. Safety critical systems are very sensitive to soft errors. A bit flip due to soft error can change the value of critical variable and consequently the system control flow can completely be changed which leads to system failure. To minimize soft error risks, a novel methodology is proposed to detect and recover from soft errors considering only 'critical code blocks' and 'critical variables' rather than considering all variables and/or blocks in the whole program. The proposed method shortens space and time overhead in comparison to existing dominant approaches.*

## 1. INTRODUCTION

In recent Times, performance of microprocessors has been increasing remarkably due to improved fabrication technology. Smaller feature sizes, lower operating voltage levels, and reduced noise margins have also helped to improve the performance and lower the power consumption of today's microprocessors. On the other hand, these advancements have made chips more prone to transient faults, which are temporary unintended changes of states resulting from the latching of single-event transient (transient voltage fluctuations); consequences of external particle strikes or process-related parametric variation. These transient faults when executed in the system create soft error. Unlike manufacturing or design error, soft error does not occur consistently rather these are transitory. Soft error involves changes to data by altering signal transfers and stored values - the charge in a storage circuit, as instance, and thereby, resulting in incorrect program execution. These do not cause permanent physical damage to the chip.

The undesired change due to these errors may alter the control flow of the system and may be catastrophic for the desired functionalities of the system. Specially, they are matters of great concern in those systems where high reliability is necessary [4], [5], [6]. Space programs (where a system cannot malfunction while in flight), banking transaction (where a momentary failure may cause a huge difference in balance), automated railway system (where a single bit flip can cause a drastic accident) [7], mission critical embedded applications, and so forth, are a few examples where soft errors are severe.

Soft errors mitigating techniques mostly focuses on post design phases *i.e.* circuit level solutions, logic level solutions, error correcting code, spatial redundancy, etc., and some software based solutions evolving duplication of the whole program or duplication of instructions [8], Critical Variable re-computation in whole program [9], etc. are concerns of prior research. Duplication seems to provide high-coverage at runtime for soft errors; it makes a comparison after every instruction leaving high performance overhead to prevent error-propagation and resulting system crashes. It compares the results of replicated instructions at selected program points such as stores and branches [4]. While this reduces the performance overhead, it sacrifices coverage as the program may crash before reaching the comparison point [9]. Error Detection by Diverse Data and Duplicated Instruction [8] is a software-based diverse execution technique in which original program is transformed into a different program where each data operand is multiplied by a constant value. The same processor executes the original program and the transformed one and the results are compared. Error Detection by Diverse Data and Duplicated Instruction can't detect errors in instruction issue and decode logic as it introduces diversity only in the data values used in the program and not in the instructions that compute the data values.

For soft errors, designers often reimburse by adding hardware redundancy and making circuit and process-level adjustments. However, different types of applications have different data integrity and availability demands, which make hardware approaches such as these too costly for many markets. In this respect, software techniques can provide fault tolerance at a lower cost and with superior flexibility since they can be selectively deployed in the field even after the hardware has been manufactured. Most existing software-only techniques use recompilation, requiring access to program source code. Regardless of the code transformation method, existing techniques also sustain unnecessary momentous performance penalties by uniformly protecting the entire program without taking into account the varying susceptibility of different program regions and state elements to soft errors. Software based approaches can significantly improve dependability without requiring hardware modifications. This criterion makes software redundancy techniques significantly cheaper and easier to deploy than hardware based approaches, working even on machines already in the field. Exploitation of redundancy techniques in the field may become important because of poor estimates of the severity of the soft-error rate by designers and because of the uncertainty in the usage condition of the machine. Changes to the operating environment of the hardware can also have a noticeable effect on reliability, requiring the deployment of software redundancy techniques.

Many static approaches [67], [68] have been proposed so far to find soft errors in programs, which have proven effective in finding errors of known types only. But there is still a large gap in providing high-coverage and low-latency (rapid) error detection to protect systems from soft error while the system is in operation. Hence, this paper proposed a new approach for soft error detection and recovery technique, which lessen time and memory overhead by working with critical blocks and variables within program code.

The paper is organized as follows: Section 2 describes related works. Section 3 outlines the methodology. Experimental Analysis is shown in Section 4. Finally in Section 5, conclusions are stated.

## 2. Related Work

A good number of works have been performed on soft errors mitigation. Three types of soft errors mitigation techniques are highlighted so far; that are 'Software based' approach, 'Hardware based' solutions and 'Hardware and software combined' approach which may be stated as 'hybrid'.

Software based approaches to tolerate soft errors include redundant programs to detect and/or recover from the problem, duplicating instructions [13], [14], task duplication [15], dual use of super scalar data paths, and Error detection and Correction Codes (ECC) [16]. Chip level Redundant Threading (CRT) [12] used a load value queue such that redundant executions can always see an identical view of memory. Walcott et al. [56] used redundant multi threading to determine the architectural vulnerability factor, and Shye et al. [57] used process level redundancy to detect soft errors. In redundant multi threading, two identical threads are executed independently over some period and the outputs of the threads are compared to verify the correctness. EDDI [54] and SWIFT [55] duplicated instructions and program data to detect soft errors. Both redundant programs and duplicating instructions create higher memory requirements and increase register pressure. Error detection and Correction Codes (ECC) [16] adds extra bits with the original bit sequence to detect error. Using ECC to combinational logic blocks is complicated, and requires additional logic and calculations with already timing critical paths.

Hardware solutions for soft errors mitigation mainly emphasize circuit level solutions, logic level solutions and architectural solutions. At the circuit level, gate sizing techniques [17], [18], [19] increasing capacitance [20], [21], resistive hardening [22] are commonly used to increase the critical charge ($Q_{crit}$) of the circuit node as high as possible. However, these techniques tend to increase power consumption and lower the speed of the circuit. Logic level solutions [58], [59], [60] mainly propose detection and recovery in combinational circuits by using redundant or self-checking circuits. Architectural solutions mainly introduce redundant hardware in the system to make the whole system more robust against soft errors. They include dynamic implementation verification architecture (DIVA) [23].

Hardware and software combined approaches [24], [25], [29], [26], [30], [27] use the parallel processing capacity of chip multiprocessors (CMPs) and redundant multi threading to detect and recover the problem. Mohamed et al. [62] shows Chip Level Redundantly Threaded Multiprocessor with Recovery (CRTR), where the basic idea is to run each program twice, as two identical threads, on a simultaneous multithreaded processor. One of the more interesting matters in the CRTR scheme is that there are certain faults from which it cannot recover. If a register value is written prior to committing an instruction, and if a fault corrupts that register after the committing of the instruction, then CRTR fails to recover from that problem. In Simultaneously and Redundantly Threaded processors with Recovery (SRTR) scheme [65], there is a probability of fault corrupting both threads since the leading thread and trailing thread execute on the same processor. Others [61], [66], [63], [64] have followed similar approaches. However, in all cases the system is vulnerable to soft error problems in key areas.

In contrast, the complex use of threads presents a difficult programming model in software-based approaches while in hardware-based approaches, duplication suffer not only from overhead due to synchronizing duplicate threads but also from inherent performance overhead due to additional hardware. Moreover, these post-functional design phase approaches can increase time delays and power overhead without offering any performance gain.

## 3. The Methodology to Mitigate Soft Errors Risks

A novel methodology has been proposed to mitigate soft error risk. In this method, the major working phenomenon consists of two-phases. During 1$^{st}$ phase, the proposed method detects soft errors at critical blocks and critical variables. At 2$^{nd}$ phase, the recovery mechanism goes to action by replacing the erroneous variable or code block with originals. The overall methodology is depicted in Figure 2.

The program that is to be executed is split into blocks; among them, 'critical blocks' are identified. Critical Blocks (CB) provide high coverage for data value errors. These include decisive programming code fragments and treated as special program-segments besides other blocks or segments. Advancing with program continuation depends on these blocks. The code blocks that determine branching of the program control flow are termed as critical-blocks. The dashed block in Figure 1 is example of CB within a program code segment. Critical blocks make decision which of the distinctive paths will be followed.

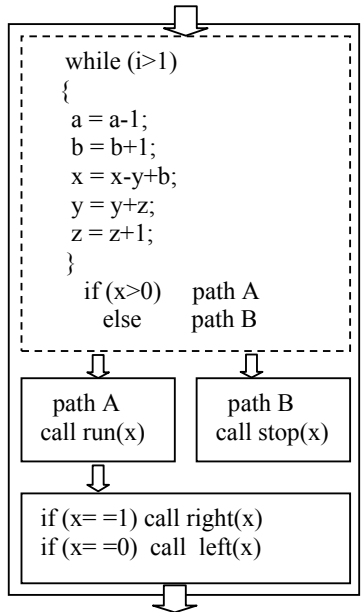

Figure 1. Critical Block

Critical variables (CV) provide high coverage for data value errors. Variables in a program that exhibit high sensitivity to random data errors in the application are critical variables. Placing checks on critical variables achieves high detection coverage. Critical variables are defined as those, which assumes great sensitivity to errors, and deriving error detectors for these variables provides great exposure for errors in any data value used in the program.

As Figure 2 illustrates, while the program is being executed, if critical blocks are encountered, they are treated specially. To meet these circumstances, a computational procedure is invoked and then the critical variables within the block are computed twice to get two outcomes. If the comparing-mechanism finds the results identical, the ordinary execution flow of the program continues from the next block. Otherwise, the variable or block is noticed to be affected by soft error.

At the $2^{nd}$ phase, the recovery process responses. The task includes replacing the found erroneous critical block or variable with the original program-block/variable. To serve the purpose, a backup of the original program is managed earlier.

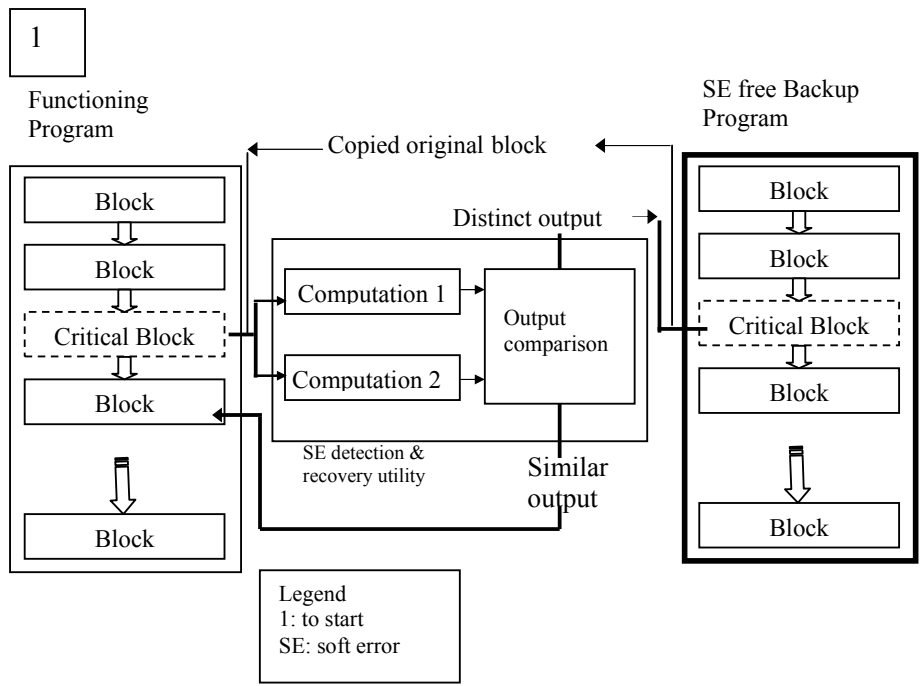

Figure 2. Soft Error Detection and Recovery

### 3.1. Creation and Storing Back up of operational program

It is the 1st and crucial step of the methodology. A backup of the examinee program is created and later on, the backup of that functioning program is kept in memory for further functioning. This backup is assumed to be soft error free. Diverse technique - ECC, Parity, RAID, and CRC or other techniques can be adopted to ensure soft error free back up.

### 3.2. Critical Variables and Blocks Identification

All variables or blocks are not equally responsible or susceptible to system failure in a computer program. These can be analyzed according to their significance level and responsibility to system mal-function or failure. Variables are of higher levels of significance are considered most vulnerable.

Critical variables (CV) and Critical Blocks (CB) provide high coverage for data value errors. By dint of CV and CB, program execution flow is determined and assuming erroneous values diversified from originals, these lead to erroneous outcome of program. Variables are determined as 'critical' through adopting and considering some phenomenon like number of recursion, dependencies, etc. Criticality ranges higher with respect to more dependencies.

```
while (i>1)
{
 a = a-1;
 b = b+1;
 x = x-y+b;
 y = y+z;
 j = j+1;
 z = z+1;
}
```

Figure 3. An Example Program-segment to show Variable Dependency

Different ways of variable dependence-relationships exist. As instance, a variable can be '*data dependent*' or '*control dependent*' on other variables. Data dependency uses pairs of program points. A variable $v$ defined at $n$ is said to be 'data dependent' on another variable $v_0$ at $n_0$ if the latter is used in $n$ and there exists an execution path from $n_0$ to $n$ on which $v_0$ is not customized. Conversely, control dependency arises from branching in program due to statements that are conditional or from function calls. A variable $v$ defined at $n$ is said to be 'control dependent' on another variable $v_0$ at $n_0$ if the execution or non-execution of $n$ depends on the truth-value of the expression involving $v_0$ at $n_0$. The program code fragment shown in Figure 3, which depicts a simple program with a single while-loop, narrates ideas of variable dependence relationships. At the end of the program, variable x depends upon the initial values of the variables i, z, y, b, x. Here, the way is shown in which variable dependence can be circular or loop carried (x's dependence upon y) and involves control dependence (x's dependence upon i) as well as data dependence (other dependences stated here).

Besides, variable dependencies may be classified as 'forward dependencies', and 'backward dependencies'.

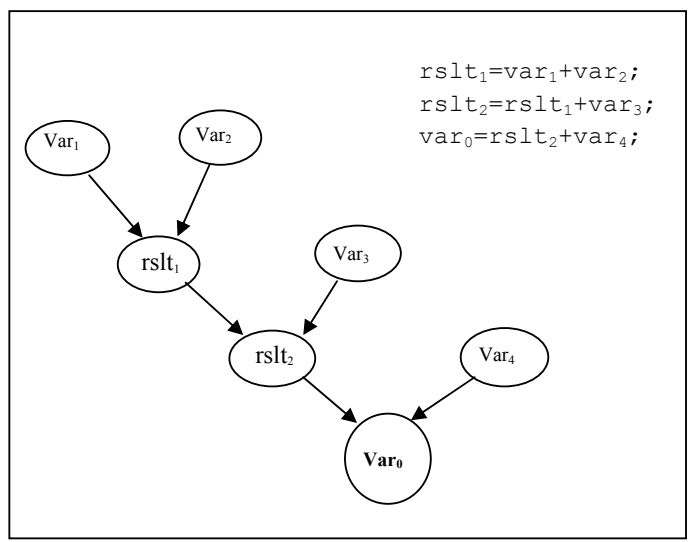

Figure 4. Backward dependency graph

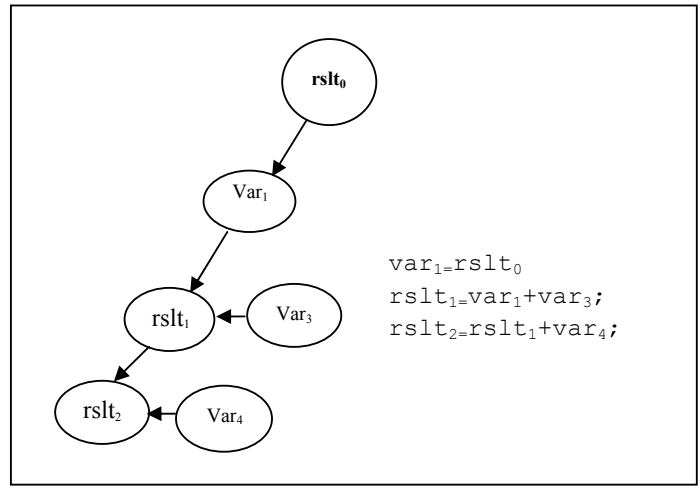

Figure 5. Forward dependency graph

Figure 4 and Figure 5 shows the variable dependency graphs for backward and forward dependencies respectively. In the figures, three statements are considered and assumed as part of a program code segment and seven variables are deployed. As shown in Figure 4, while executing, value determination of variable 'rslt$_2$' of statement 2 will depend on statement 1 for result of the variable 'rslt$_1$' , statement 3 will depend on statement 2 to calculate 'var$_0$' since 'var$_0$'is the summation of 'rslt$_2$'and 'var$_4$'. Hence, 'var$_0$'is dependent on the variables at the back e.g., 'rslt$_2$', 'var$_4$', 'rslt1' etc. This is called 'back ward dependency'. Any error in rslt$_2$, var$_1$, rslt$_1$ etc. will be propagated to var$_0$. If soft errors occur in any of the variables, it can be detected by comparing and checking only var$_0$.

As shown in Figure 5, statement 2 and statement 3 are dependent on 'rslt$_0$' in statement 1. Hence, the variables at the forward e.g., rslt$_2$, rslt$_1$ etc. are dependent on 'rslt$_0$'. This dependency is called 'forward dependency'. Considering the assignment statements, the tree in Figure 5 is formed and the root node (rslt$_0$) is determined. It is seen that the root node is more critical among others because it is (root node) decisive in those respect. If soft errors occurred in any node other than root, that will ultimately be propagated to the root node.

So, to detect soft errors, the critical variable comparison will be more efficient rather than consider all variables to be compared. This significantly reduces the execution time of program as well as memory utilization. Thus it may considerably increase the efficiency of error detection process.

Critical Blocks (CB) are programming segments; the program control flow depends highly on these blocks. Identification of these blocks and advancing with the critical blocks and/or variables noticed within it, is the key concept of the proposed method.

The code blocks responsible for branching the program control flow are recognized to be critical-blocks.

### 3.3 Soft Error Detection and Recovery

The critical variables found in specific critical block are computed twice each and then compared two distinct outcome to determine whether they are identical or not. As Figure 1, while executing a full program code, faced critical block is executed twice and the result is compared. If the values found are identical then program execution flow continues to the next (next block). If the result of the two computations is distinct, the total block of program is marked and replaced by the relevant program code block that is in original. And program execution is continued from the current code block.

The basic steps can be stated as follows:

**Step 1:** Each critical block is recomputed. It is executed twice and received outcomes are stored individually.

**Step 2:** Recomputed results are then compared to make sure that they are identical / consistent.

**Step 3:** If recomputed values show consistency, program will be continued from the next code block and no soft error will be reported.

**Step 4:** If recomputed values show inconsistency, particular program block will be identified as erroneous and soft error will be reported; and then the recovery procedure will be called for. With the block replacing procedure, the significant erroneous critical block will be replaced by the relevant original program's critical block. And the program execution will be continued from current block.

## 4. Experimental Analysis

The methodology is experimented through a multi phases simulation process. The simulation procedure mainly evolves 'error-detection' phase and 'error-recovery' phase. It detects the soft error occurred through the detection phase and duly recovers it in order to lead the program towards expected output with its counterpart; the recovery process. Through a backup creation phase, a backup of the operational program is created and kept in memory for soft error recovery process. A candidate program is checked through the simulator to detect soft error and duly reports it if there is any. Block wise execution of core program along with twin computation of critical one is helped by this backup to go to the desired end by supporting the blocks to be actual valued. A binary representation of the executable program is formed and lunched on the simulator editor as hexadecimal format. They are sequence of bit stream to negotiate with. Fault is injected manually that is flipping a bit/ bits to change the original code sequence.

### 4.1 Injection of Faults

In 'fault injection' phase, fault is injected manually through a fault-injection wizard of the simulator. Fault injection evolves bit flipping; this is to change a binary '0' (zero) to a binary '1' (one) and vice-versa at any bit position of a particular byte. Fault is injected at variables and/ or at any random position (instructions or variables) of the program's binary file to change the value of the variables or the instructions. If binary representation of a variable's value is 01000110, bit flipping may occur at any bit position due to error. If bit position 5 (say) will be flipped from 0 to 1, the value of the variable will be 01001110 what will cause a huge difference for value of the variable. It is done so that in detection-phase, how the mechanism treats the fact - whether any crucial change on output of the program takes place due to injected fault(s) and that is detected; or minor change occurs having less significance to output of the program and detector ignores it treating it as benign fault(s) that is the program may run and produce a relatively realistic output.

## 4.2 Detection of Soft Errors

After fault-injection at random position of the candidate program's new representation as shown in following figure, detection process is to go ahead. The critical blocks come into main focus among all other program-blocks. While executing, they are computed twice and compared by values they contain. The error detection tool detects the mismatch(es) as error (as shown in Figure 6) if the comparison is distinguished-valued.

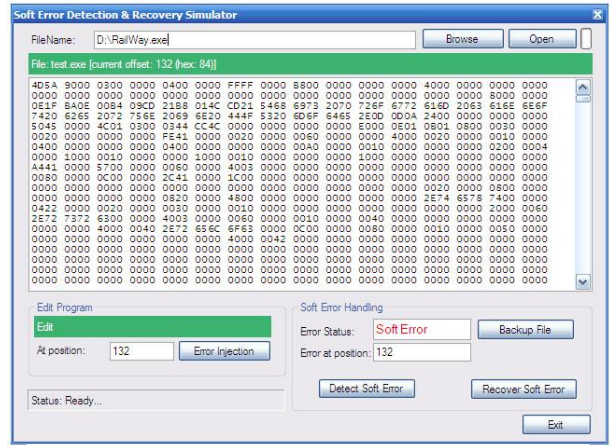

Figure 6. Simulator interface

As soft error is transitory, it has neither repetitive occurrence nor long lasting effect. Hence, consecutive computation of variables results different result if any of them assumes erroneous value that should not to be. If no such case is encountered, 'no soft error' is reported by the utility.

## 4.3 Recovery from Soft Errors

In recovery phase, the previously noticed erroneous critical variable and its location are traced; and then, the backup is invoked to replace the erroneous critical code block with relevant originals. Soft error recovery tool in the interface activates the mechanism to perform recovery process.

## 4.4 Result Analysis

The proposed methodology has gone through an experimental simulation process. It exhibits some optimistic result-oriented outcomes.

The methodology is found to have to deal with some constant amounts of critical variables in recognized critical blocks other than to go through considering all the variables in the critical system program code. This makes it less time consuming to perform computation as shown in Figure 7.

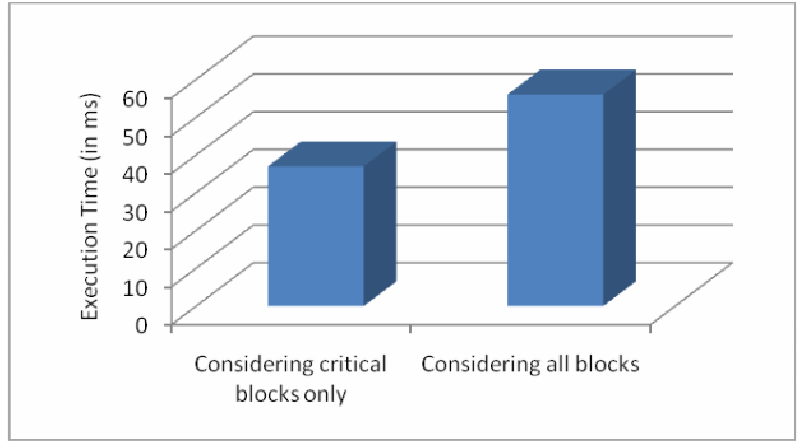

Figure 7. Comparison of Execution Time

In contrast, treating with fewer variables (only in critical blocks) will definitely require some lower memory space (as shown in Figure 8) and hence, the overall program execution time will be reduced significantly. More over, if the critical variables-recomputed values are stored in L1 cache, memory access latency will be a great issue to be noticed.

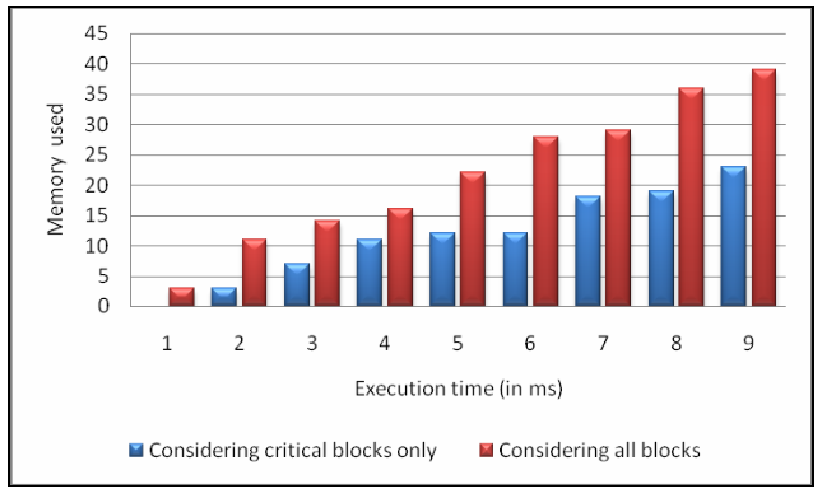

Figure 8. Comparison of memory utilization

Some other well-established methods like [8], [9], [11], [33] etc. takes longer execution time as their schemes required. Table 1 depicts a theoretical comparison among likely hood methods:

Table 1. Review of different soft-error tolerance techniques

| Approaches | Diverse Data and Duplicated Instruction ED$^4$I [8] | Critical Variables Re-computation for Transient Error Detection [9] | Source Code Modification [11], [33]] | Proposed Method |
|---|---|---|---|---|
| **Adopted methodology** | The original program and the transformed program are both executed on the same processor and the results are compared. | Recomputes only critical variables (not the instructions) to detect and recovery from soft errors. | Based on modifications of the source code. Protection methods are applied at the intermediate representation level of the compiled source code. | A backup is kept in a memory for error correction. Detection is performed by only re-computation of critical blocks. Erroneous blocks are replaced by the relevant backed up program blocks. |
| **Memory space overhead** | At least double | Depends on no. of Critical variable. | Larger than usual depending upon Modification scheme | Depends on no. of Critical Blocks |
| **Execution time overhead** | Longer than usual since comparison | Longer than usual since CV re-computation | High running time since source code modification | Relatively much lower |
| **Number of variables to be executed** | Double | Depends on no. of Critical variable. | Depends on modification techniques | Depends on no. of Critical Blocks |
| **Drawback** | Program may crash before reaching the comparison point. Possibility to be erroneous for both of the programs. | Is not able to detect all severe errors since it does work only with Critical variables. Instructions may also be erroneous. | System may crush before reaching the control flow checking point. | Improper identification of critical blocks leads to inefficiency. |

## 5. Conclusions

As soft errors become more widespread across a wide range of markets now a days, techniques, which can modify the level of protection to each user's specific performance and reliability requirements, will be needed. The possibility of modification of existing methodologies to lower the criticalities of program blocks to minimize the risk of soft errors is reflected here. The significant contribution of the proposed method is to lower soft error risks with a minimum time and space complexity since it works only with critical variables in critical blocks; hence, all the variables in program code are not considered to be recomputed or replicated though they also may cause such error, that is from benign faults and faults that are not severe for the system; which does no interference to system performance. It is seen that only critical variables induced soft error affects systems program flow in a great extent to be malfunctioned. Hence, leaving some ineffective errors un-pursued, the proposed method can achieve the goal in a cost effective (with respect to time and space) way.

Some possible steps could be adopted to enhance the performance of the method. Storing the recomputed outcomes of the critical blocks at cache memory will enhance memory access time, which can be a significant issue in case of memory latency. This can make the proposed method less time consuming by compensating the time killing to make double computation. The protection of backup of original program is a great concern to remain it soft error free. In support of storage, besides existing techniques such as Error-Correcting-Code (ECC), Redundant Array of Inexpensive Disks (RAID), enhancement can be explored. Another considerable concern is the critical blocks and critical variables. Obviously critical blocks and critical variables identification among numerous code blocks in operating program is a great challenge to make it optimum. Efficiency of the proposed method is mostly depends on proper identification of critical blocks and critical variables. They can be grouped according to their criticality that is treated differently in different view-points. Several issues like "*fan out*", that is number of dependency/ branches exist; number of "*recursion*"- that is, looping or how many times a call repeated; "*severity of blocks*", that is block containing more weighted variables etc., are wide open to determine the criticality. Hence, much more scopes are available in the field of critical block and variable identification. In contrast, proper identification of these critical blocks and variables and work only with them can mitigate most of the soft error risks with time and space optimality.

#### ACKNOWLEDGEMENTS

The authors would like to thank Bishnu Sarker, Tanay Roy, Department of Computer Science & Engineering, Khulna University of Engineering & Technology, Khulna-9203, Bangladesh for their structural comments and valuable supports.

#### REFERENCES

[1]  A. Timor, A. Mendelson, Y. Birk, and N. Suri, "Using under utilized CPU resources to enhance its reliability," Dependable and Secure Computing, IEEE Transactions on, vol. 7, no. 1, pp. 94-109, 2010.

[2]  E. L. Rhod, C. A. L. Lisboa, L. Carro, M. S. Reorda, and M. Violante, "Hardware and Software Transparency in the Protection of Programs Against SEUs and SETs," Journal of Electronic Testing, vol. 24, pp. 45-56, 2008.

[3]  S. S. Mukherjee, J. Emer, and S. K. Reinhardt, "The soft error problem: an architectural

perspective," in 11th International Symposium on High-Performance Computer Architecture, San Francisco, CA, USA, pp. 243 - 247, 2005, pp. 243-7.

[4] R. K. Iyer, N. M. Nakka, Z. T. Kalbarczyk, and S. Mitra, "Recent advances and new avenues in hardware-level reliability support,"Micro, IEEE, vol. 25, pp. 18-29, 2005.

[5] V. Narayanan and Y. Xie, "Reliability concerns in embedded system designs,"Computer, vol. 39, pp. 118-120, 2006.

[6] S. Tosun, "Reliability-centric system design for embedded systems," Ph.D. Thesis, Syracuse University, United States --New York, 2005.

[7] Muhammad Sheikh Sadi, D. G. Myers, Cesar Ortega Sanchez, and Jan Jurjens, "Component Criticality Analysis to Minimizing Soft Errors Risk." Comput Syst Sci & Eng (2010), vol 26 no 1 September 2010.

[8] Nahmsuk Oh, Subhasis Mitra, Edward j. McClusky, "ED$^4$I: Error Detection by Diverse Data and Duplicated Instructions." EEE Transactions on Computers Vol. 51 No. 2 February 2002

[9] Karthik Pattabiraman, Zbigniew Kalbarczyk, and Ravishankar K. Iyer, "Critical Variable Recomputation for Transient Error Detection", 2008

[10] Chen Cuiting, "A New Hybrid Fault Detection Technique for Systems-on-a-Chip", July 1, 2008

[11] Adam Piotrowski, Dariusz Makowski, Grzegorz Jabło´nski, Andrzej, Napieralski, "The Automatic Implementation of Software Implemented Hardware Fault Tolerance Algorithms as a Radiation-Induced Soft Errors Mitigation Technique" Nuclear Science Symposium Conference Record, IEEE, 2008

[12] S. S. Mukherjee, M. Kontz, and S. K. Reinhardt, "Detailed design and evaluation of redundant multi-threading alternatives," in 29th Annual International Symposium on Computer Architecture, pp. 99-110, 2002, pp. 99-110.

[13] N. Oh, P. P. Shirvani, and E. J. McCluskey, "Error detection by duplicated instructions in super-scalar processors," Reliability, IEEE Transactions on, vol. 51, pp. 63-75, 2002.

[14] G. A. Reis, J. Chang, N. Vachharajani, R. Rangan, and D. I. August, "SWIFT: software implemented fault tolerance," Los Alamitos, CA, USA, 2005, pp. 243-54.

[15] Y. Xie, L. Li, M. Kandemir, N. Vijaykrishnan, and M. J. Irwin, "Reliability-aware co-synthesis for embedded systems," in 15th IEEE International Conference on Application-Specific Systems, Architectures and Processors, 2004, pp. 41-50.

[16] C. L. Chen and M. Y. Hsiao, "Error-Correcting Codes for Semiconductor Memory Applications: A State-Of-The-Art Review," IBM Journal of Research and Development, vol. 28, pp. 124-134, 1984.

[17] J. K. Park and J. T. Kim, "A soft error mitigation technique for constrained gate-level designs," IEICE Electronics Express, vol. 5, pp. 698-704, 2008.

[18] N. Miskov-Zivanov and D. Marculescu, "MARS-C: modeling and reduction of soft errors in combinational circuits," Piscataway, NJ, USA, 2006, pp. 767-72.

[19] Z. Quming and K. Mohanram, "Cost-effective radiation hardening technique for combinational logic," Piscataway, NJ, USA, 2004, pp. 100-6.

[20] Oma, M. a, D. Rossi, and C. Metra, "Novel Transient Fault Hardened Static Latch," Charlotte, NC, United states, 2003, pp. 886-892.

[21] P. R. STMicroelectronics Release, "New chip technology from STmicroelectronics eliminates soft error threat to electronic systems," Available at www.st.com/stonline/press/news/year2003/t1394h.htm, 2003.

[22]     L. R. Rockett, Jr., "Simulated SEU hardened scaled CMOS SRAM cell design using gated resistors," IEEE Transactions on Nuclear Science, vol. 39, pp. 1532-41, 1992.

[23]     T. M. Austin, "DIVA: a reliable substrate for deep submicron microarchitecture design," in 32nd Annual International Symposium on Microarchitecture, 1999, pp. 196-207.

[24]     B. T. Gold, J. Kim, J. C. Smolens, E. S. Chung, V. Liaskovitis, E. Nurvitadhi, B. Falsafi, J. C. Hoe, and A. G. Nowatzyk, "TRUSS: a reliable, scalable server architecture,"Micro, IEEE, vol. 25, pp. 51-59, 2005.

[25]     S. Krishnamohan, "Efficient techniques for modeling and mitigation of soft errors in nanometer-scale static CMOS logic circuits," Ph.D. Thesis, Michigan State University, United States -- Michigan, 2005.

[26]     A. G. Mohamed, S. Chad, T. N. Vijaykumar, and P. Irith, "Transient-fault recovery for chip multiprocessors," IEEE Micro, vol. 23, p. 76, 2003.

[27]     J. Srinivasan, S. V. Adve, P. Bose, and J. A. Rivers, "The case for lifetime reliability-aware microprocessors," in 31st Annual International Symposium on Computer Architecture, 2004, pp. 276-287.

[28]     M. W. Rashid, E. J. Tan, M. C. Huang, and D. H. Albonesi, "Power-efficient error tolerance in chip multiprocessors," Micro, IEEE, vol. 25, pp. 60-70, 2005.

[29]     T. N. Vijaykumar, I. Pomeranz, and K. Cheng, "Transient-fault recovery using simultaneous multithreading," in 29$^{th}$ Annual International Symposium on Computer Architecture, 2002, pp. 87-98.

[30]     Y. Xie, L. Li, M. Kandemir, N. Vijaykrishnan, and M. J. Irwin, "Reliability-aware co-synthesis for embedded systems," in 15th IEEE International Conference on Application-Specific Systems, Architectures and Processors, 2004, pp. 41-50.

[31]     S. M. Seyed-Hosseini, N. Safaei, and M. J. Asgharpour, "Reprioritization of failures in a system failure mode and effects analysis by decision making trial and evaluation laboratory technique," Reliability Engineering & System Safety, vol. 91, pp. 872-81, 2006.

[32]     S. Krishnamohan, "Efficient techniques for modeling and mitigation of soft errors in nanometer-scale static CMOS logic circuits," Ph.D. Thesis, Michigan State University, United States -- Michigan, 2005.

[33]     Adam Piotrowski, Szymon Tarnowski, "Compiler-level Implementation of Single Event Upset Errors Mitigation Algorithms."

[34]     R. C. Baumann, "Soft errors in advanced semiconductor devices — part I: the three radiation sources," Device and Materials Reliability, IEEE Transactions on Volume 1, Issue 1, 2001.

[35]     "Radiation-induced soft errors in advanced semiconductor technologies," IEEE Transactions on Device and Materials Reliability, Vol. 5, No. 3,, 2005.

[36]     M. Rebaudengo and M. Sonza Reorda and M. Violante, "A new approach to software-implemented fault tolerance," IEEE Latin American Test Workshop, vol. 40, pp. 433–437, 2002.

[37]     O. Goloubeva and M. Rebaudengo and M. Sonza Reorda and M. Violante, "Soft-error detection using control flow assertions," 18th IEEE International Symposium on Defect and Fault Tolerance in VLSI Systems (DFT'03), p. 581, 2003.

[38]     N. Oh, "Software implemented fault tolerance," Ph.D. dissertation, Stanford University, 2000.

[39]     George A. Reis, Jonathan Chang, Neil Vachharajani, Ram Rangan, David I. August, Shubhendu S. Mukherjee, "Design and Evaluation of Hybrid Fault-Detection Systems",

IEEE 2005

[40] L. L. Pullum, Software fault tolerance techniques and implementation. Artech House, Inc., 2001.

[41] S.E. Michalak and K.W. Harris and N.W. Hengartner and B.E. Takala and S.A. Wender,, "Predicting the number of fatal soft errors in los alamos national laboratory's asc q supercomputer," IEEE Transactions on Device and Materials Reliability, 2005.

[42] F. L. Kastensmidt and L. Carro and R. Reis, Fault-Tolerance Techniques for SRAM-Based FPGAs. Springer Science+Business Media, LLC, 2006.

[43] M. Rebaudengo and M. Reorda and M. Torchiano and M. Violante, "Soft-error detection through software fault-tolerance techniques," IEEE DFT'99: IEEE International Symposium on Defect and Fault Tolerance in VLSI Systems, pp. 210–218, 1999.

[44] A. Piotrowski and D. Makowski and S. Tarnowski and A. Napieralski, "Automatic implementation of radiation protection algorithms in programs generated by gcc compiler," EPAC 2008 - European Particle Accelerator Conference, 23-27 June, Genoa (Italy), 2008.

[45] N.S. Oh, P.P. Shirvani and E.J. McCluskey, Error detection by duplicated instructions in super-scalar processors. IEEE Transactions on Reliability, 51(1):63-- 75, March 2002.

[46] N.S. Oh, S. Mitra and E.J. McCluskey, ED$^4$I: Error Detection by diverse data and duplicated instructions in super-scalar processors. IEEE Transactions on Reliability, 51(2): pp. 180-199, February 2002.

[47] S. M. Yacoub and H. H. Ammar, "A methodology for architecture-level reliability risk analysis," IEEE Transactions on Software Engineering, vol. 28, pp. 529-547, 2002.

[48] A. G. Mohamed, S. Chad, T. N. Vijaykumar, and P. Irith, "Transient-fault recovery for chip multiprocessors," IEEE Micro, vol. 23, pp. 76, 2003.

[49] H. T. Nguyen, Y. Yagil, N. Seifert, and M. Reitsma, "Chip-level soft error estimation method," Device and Materials Reliability, IEEE Transactions on, vol. 5, pp. 365-381, 2005.

[50] Maurizio Rebaudengo, Matteo Sonza Reorda, Marco Torchiano, "A source-to-source compiler for generating dependable software."

[51] F. Li, G. Chen, M. Kandemir, I. Kolcu, "Improving Scratch-Pad Memory Reliability Through Compiler-Guided Data Block Duplication.", IEEE 2005.

[52] B. Nicolescu and R. Velazco, "Detecting soft errors by a purely software approach: method, tools and experimental results," in Proc. of Design, Automation and Test in Europe Conference and Exhibition, Munich, Germany, Mar. 2003.

[53] P. P. Shirvani, N. Saxena, and E. J. McCluskey, "Software-implemented edac protection against SEUs," IEEE Transaction on Reliability, vol. 49, no. 3, pp. 273–284, Sept. 2000.

[54] N. Oh, P. P. Shirvani, and E. J. McCluskey, "Error detection by duplicated instructions in super-scalar processors, "*Reliability, IEEE Transactions on,* vol. 51, pp. 63-75, 2002.

[55] G. A. Reis, J. Chang, N. Vachharajani, R. Rangan, and D. I. August, "SWIFT: software implemented fault tolerance," Los Alamitos, CA, USA, 2005, pp. 243-54.

[56] K. R. Walcott, G. Humphreys, and S. Gurumurthi, "Dynamic prediction of architectural vulnerability from microarchitectural state," New York, NY 10016-5997, United States, 2007, pp. 516-527.

[57] A. Shye, J. Blomstedt, T. Moseley, V. Janapa Reddi, and D. Connors, "PLR: A Software Approach to Transient Fault Tolerance for Multi-Core Architectures,"*Dependable and Secure Computing, IEEE Transactions on,* To be Appeared.

[58] M. Z. S. Mitra, N. Seifert, TM Mak and K. Kim. Soft and IFIP, "Soft Error Resilient System Design through Error Correction,"*VLSI-SoC,* January, 2006.

[59] M. Zhang, "Analysis and design of soft-error tolerant circuits," Ph.D. Thesis, University of Illinois at Urbana-Champaign, United States -- Illinois, 2006.

[60] M. Zhang, S. Mitra, T. M. Mak, N. Seifert, N. J. Wang, Q. Shi, K. S. Kim, N. R. Shanbhag, and S. J. Patel, "Sequential Element Design With Built-In Soft Error Resilience," *Very Large Scale Integration (VLSI) Systems, IEEE Transactions on,* vol. 14, pp. 1368-1378, 2006.

[61] S. Krishnamohan, "Efficient techniques for modeling and mitigation of soft errors in nanometer-scale static CMOS logic circuits," Ph.D. Thesis, Michigan State University, United States -- Michigan, 2005.

[62] A. G. Mohamed, S. Chad, T. N. Vijaykumar, and P. Irith, "Transient-fault recovery for chip multiprocessors," *IEEE Micro,* vol. 23, p. 76, 2003.

[63] J. Srinivasan, S. V. Adve, P. Bose, and J. A. Rivers, "The case for lifetime reliability-aware microprocessors," in *31st Annual International Symposium on Computer Architecture,* 2004, pp. 276-287.

[64] M. W. Rashid, E. J. Tan, M. C. Huang, and D. H. Albonesi, "Power-efficient error tolerance in chip multiprocessors," *Micro, IEEE,* vol. 25, pp. 60-70, 2005.

[65] T. N. Vijaykumar, I. Pomeranz, and K. Cheng, "Transient-fault recovery using simultaneous multithreading," in *29$^{th}$ Annual International Symposium on Computer Architecture,* 2002, pp. 87-98.

[66] Y. Xie, L. Li, M. Kandemir, N. Vijaykrishnan, and M. J. Irwin, "Reliability-aware co-synthesis for embedded systems," in *15th IEEE International Conference on Application-Specific Systems, Architectures and Processors,* 2004, pp. 41-50.

[67] D. Evans, J. Guttag, J. Horning, and Y. M. Tan. *LCLint: A tool for using specifications to check code*. In Proc. Symposium on the Foundations of Software Engineering (FSE), Dec.1994.

[68] Michael D. Ernst, Jake Cockrell, William G. Griswold, and David Notkin. *Dynamically discovering likely program invariants to support program evolution*. IEEE Transactions on Software Engineering, 27(2):1-25, 2001.

**Authors**

Short Biography

Muhammad Sheikh Sadi received B.Sc. Eng. in Electrical and Electronic Engineering from Khulna University of Engineering and Technology, Bangladesh in 2000, M.Sc. Eng. in Computer Science and Engineering from Bangladesh University of Engineering and Technology, Dhaka, Bangladesh in 2004, and completed PhD (Area: Dependable Embedded Systems) from Curtin University of Technology, Australia in 2010. He is currently Associate Professor at the Department of Computer Science and Engineering, Khulna University of Engineering and Technology, Bangladesh. He teaches and supervises undergraduate and postgraduate theses in topics related to Embedded Systems, Digital System Design, Soft Errors Tolerance etc. He has published over 20 papers and book chapters in his area of expertise. Muhammad Sheikh Sadi is a member of the IEEE since 2004.

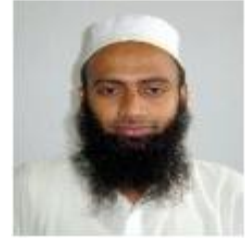

Md. Mizanur Rahman Khan received his B.Sc.(Hons) degree in Computer Science from National University, Bangladesh in 2007. He is the Associate Member of Bangladesh Computer Society (BCS). He is currently a student of M.Sc. Eng. at the Department of Computer Science and Engineering, Khulna University of Engineering and Technology (KUET), Bangladesh. He has published a paper in the field of Soft Errors Tolerance and written a paper on anti-phishing (yet to be published) as well. His current research interests are in Embedded Systems, Soft Errors Tolerance and System Modelling.

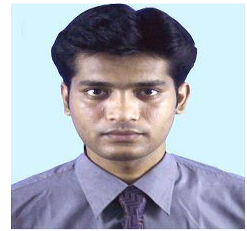

Md. Nazim Uddin received B.Sc. Eng. in Computer Science and Engineering from Khulna University of Engineering and Technology, Bangladesh in 2007. He is the Member of The Institute of Engineers, Bangladesh (IEB) and Associate Member of Bangladesh Computer Society (BCS). He is currently a student of M.Sc. Engg. in the Department of Computer Science and Engineering, Khulna University of Engineering and Technology, Bangladesh. He has also written a peer-reviewed paper in the field of Soft Errors Tolerance. His current research interests are in Embedded Systems, Soft Errors Tolerance and System Modeling.

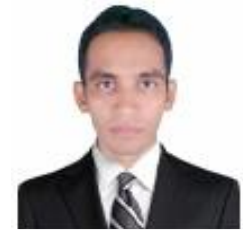

Jan Jürjens is a Professor at the Chair for Software Engineering in the Department of Computer Science of Technical University Dortmund (Germany), the Scientific Coordinator "Enterprise Engineering" and Attract research group leader at the Fraunhofer Institute for Software and Systems Engineering ISST (Dortmund), and a Senior Member of Robinson College (Univ. Cambridge, UK). He obtained his Doctor of Philosophy in Computing from the University of Oxford and author of "Secure Systems Development with UML" (Springer, 2005; Chinese translation: Tsinghua University Press, Beijing, 2009) and various publications mostly on computer security and software engineering, totaling more than 2000 citations (Google Scholar, Apr. 2010). Much of his work is done in cooperation with industrial partners including Microsoft Research (Cambridge), O2 (Germany), BMW, HypoVereinsbank, Infineon, Deutsche Telekom, Munich Re, IBM-Rational, Deutsche Bank, Allianz.

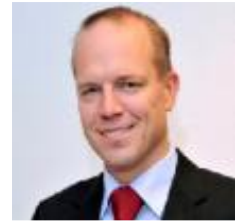